\documentstyle[preprint,aps,epsfig]{revtex}

\begin{document}
\title{Evolution of the universality class in slightly diluted ($1>p>0.8$) Ising systems.}
\author{Manuel I. Marqu\'es and Julio A. Gonzalo}
\address{Departamento de F\'isica de Materiales, C-IV
Universidad Aut\'onoma de Madrid, 28049, Madrid, Spain
email: julio.gonzalo@uam.es}
\date{14-03-2000}
\maketitle

\begin{abstract}
The crossover of a pure (undiluted) Ising system (spin per site probability $p=1$) to a diluted Ising
system (spin per site probability $p<0.8$) is studied by means of Monte Carlo calculations 
with $p$ ranging between 1 and 0.8 at intervals of 0.025. The evolution of the self averaging is 
analyzed by direct determination of the normalized square widths $R_{M}$ and $R_{\chi}$ as a function of $p$. We find a monotonous and smooth evolution 
from the pure to the randomly diluted universality class. The $p$-dependent transition 
is found to be indepent of size ($L$). This property is very convenient for extrapolation towards the
randomly diluted universality class avoiding complications resulting from  finite size effects.
\end{abstract}

Systems with quenched randomness have been studied intensively for several
decades \cite{Domb}. One of the first results was establishing the so called
Harris criterion \cite{Harris}, which predicts that a weak dilution does not
change the critical behavior's character near second order phase transitions
for systems of dimension $d$ with specific heat exponent lower than zero
(the so called P systems), $\alpha _{pure}<0\Longleftrightarrow \nu
_{pure}>2/d$ due to the hyperscaling relationship, in the undiluted case.
This criterion has been confirmed by several renormalization group (RG)
analyses \cite{Lubensky,LubenskyII,Grinstein}, and by scaling analysis \cite
{Aharony}. It was shown to hold also in strongly diluted systems by Chayes
et al. \cite{Chayes}. For $\alpha _{pure}>0$ (the so called R systems), for
example the Ising 3D case, the system fixed point flows from that of a pure
(undiluted) fixed point towards a new stable fixed point at which $\alpha
_{random}<0$ \cite{Lubensky,LubenskyII,Grinstein,Aharony,Chayes,Kinzel} for
diluted systems.

Recently Ballesteros et al. have used the Monte Carlo approach to study
diluted Ising systems in two\cite{BallesterosI}, three \cite{BallesterosII}
and four dimensions \cite{BallesterosIII}. The existence of a new
universality class for the randomly diluted Ising system (RDIS) (different
from that of the pure Ising model, and $p$-independent being $p$ the spin
per site probability) is proved using an infinite volume extrapolation
technique\cite{BallesterosII} based upon the leading correction to scaling.
The critical exponents obtained this way agree with the experimental
critical exponents for a random disposition of vacancies in diluted magnetic
systems \cite{Folk}.

The crossover from the pure Ising system $p=1$ to the randomly diluted
system may occur for very large values of the average density of occupied
sites ($1>p>0.8$), i.e. systems with a very small amount of vacancies. In
this region, the specific heat critical exponent must flow from a value
grater than zero, $\alpha _{pure}>0$ for $p=1$, to a value smaller than
zero, $\alpha _{random}<0$ for $p=0.8$. It means that in principle is
possible to expect the existence of a critical density $p_{c}$ at which its
value is equal to zero. The $p_{c}$ value has been found to be around $0.9$ 
\cite{BallesterosII}.

In principle it is not clear whether this crossover should occur smoothly or
whether the crossover should take place sharply at a critical value $p_{c},$
separating the two distinct universality classes. This is a crucial question
to establish whether the slightly diluted systems should be considered as
pure (basically undiluted), as randomly diluted systems, or, on the
contrary, as intermediate states between both extreme classes. There is an
intrinsic difficulty in detecting the evolution of critical exponents from
pure to diluted random Ising systems due to the fact that they are very
similar (see Table I). Following Ballesteros et al. we find $\alpha
_{random}=-0.051,\beta _{random}=0.3546,\gamma _{random}=1.342$ in
comparison with the pure undiluted values: $\alpha _{pure}=0.11,\beta
_{pure}=0.3267,\gamma _{{pure}}=1.237$ \cite{Blote} (incidentally, this does
not happen if the disorder is long range correlated \cite
{Weinrib,BallesterosIV,Marques}). That is why it is useful to study some
other universal quantity which clearly indicates the difference between the
pure and the random universality classes.

For a random hypercubic sample of linear dimension $L$ and number of sites $%
N=L^{d}$, any observable singular property $X$ presents different values for
different random realizations corresponding to the same average dilution.
This means that X behaves as a stochastic variable with average $[X]$,
variance $(\Delta X)^{2}$ and a normalized square width $R_{X}=(\Delta
X)^{2}/[X]^{2}$. This quantity allows us to determine properly the evolution
from the pure to the randomly diluted system by an investigation of its self
averaging behavior. A system is said to exhibit self averaging (SA) if $%
R_{X}\rightarrow 0$ as $L\rightarrow \infty $. If the system is away from
criticality, $L>>\xi $ (being $\xi $ the correlation length) the central
limit theorem indicates that strong SA must be expected. However, the
self-averaging behavior of a ferromagnet at criticality (where $\xi >>L)$ is
not so obvious. This point has been studied recently. Wiseman and Domany
(WD) have investigated the self-averaging of diluted ferromagnets at
criticality by means of finite-size scaling calculations\cite{Wiseman},
concluding weak SA for both the P and R cases. In contrast Aharony and
Harris (AH), using a renormalization group analysis in $d=4-\varepsilon $
dimensions, proved the expectation of a rigorous absence of self-averaging
in critically random ferromagnets \cite{AharonyII}. More recently, Monte
Carlo simulations where used to check this lack of self-averaging in
critically disordered magnetic systems \cite
{BallesterosII,WisemanII,AharonyIII}. The absence of self-averaging was
confirmed. The source of the discrepancy with previous scaling analysis by
WD was attributed to the particular size ($L$) dependence of the
distribution of pseudocritical temperatures used in their work.

In the present work we study the evolution of the normalized square width
from $p=1$ to $p=0.8$ at small steps $\Delta p=0.025,$ in an effort to
characterize in detail the evolution of the normalized square width from $%
R_{M}=R\chi =0$ to the zone where lack of self-averaging $(R_{M}\neq 0$ and $%
R\chi \neq 0)$ appears. We will determine whether there is some sharp
critical value $p_{c}$ separating both universality classes, or there is a
smooth evolution. We have performed Monte Carlo calculations using the Wolff
single cluster algorithm \cite{Wolff} in diluted three dimensional Ising
systems at criticality for different values of the site occupation spin
probability $1>p>0.8.$ In order to obtain good enough statistics in our
determination of the normalized square width for magnetization and
susceptibility, we have used 500 samples for the sizes $L=20,40,60.$ The
magnetization and susceptibility of each sample was determined using 50.000
MCS leaving the previous 100.000 MCS for thermalization. The critical
temperature for each dilution was taken by interpolation between the data
reported by Heuer et al. \cite{Heuer} and Ballesteros et al. \cite
{BallesterosII}. We may note that there are no much data in the literature
about the critical temperature in this region of slightly diluted systems. A
''nearly-linear'' extrapolation of the data for $T_{c}$ vs. $p$ from $p=1$
to $p=0.8$ seems clear from Fig.1. To check this we have calculated the
critical temperature for several values of $\ p$ by means of statistics on
the Binder Cumulant, and we have found that the data lie over the
interpolation functions previously considered.

We can build histograms with the values obtained for susceptibility or
magnetization at criticality. In the case of very high $p$ values,
corresponding to very-low dilution, the width of these histograms is very
small, indicating the existence of proper self-averaging. However, for
somewhat lower values of $p$, the system starts its crossover to the
randomly diluted behavior and the width of the histograms begins to
increase, indicating that lack of self-averaging is taken place. Fig.2a and
b show the evolution of these histograms for the case of the susceptibility
and for $L=60$. Note that the width of the histograms increases monotonously
as $p$ decreases, indicating a smooth flow towards the random diluted
universality class. We will see this point more clearly studying the value
of the normalized square width.

The results obtained for the normalized square width of the magnetization
and the susceptibility are presented in Fig3 and Fig4 respectively. Note how
in both cases we find a smooth evolution indicating that the crossover from
the pure fixed point to the randomly diluted fixed point takes place
smoothly and continuously and that there is no apparent critical value of $%
p_{c}$ acting as a boundary between the two regimens.

The value of the normalized square width for a given $p$, can be strongly
affected by finite size effects. In order to obtain a value of $R_{M}$ or $%
R_{\chi }$ independent of $p$ for $p<0.8$ it is necessary to consider very
high values of $L$ \cite{AharonyIII} or to use the so called infinite volume
extrapolation \cite{BallesterosII}. However, for small dilution $(1>p>0.8)$
the values of the normalized square width are nearly unaffected by finite
size effects \cite{BallesterosII}, {\bf but} they are dependent of $p$. That
is the reason why in Fig3 and Fig4 all the data seem to collapse over the
same curve. This does not happen for $p<0.8$ where finite size effects
clearly appear. To show this, we present in Fig5 data for the susceptibility
together with data by Ballesteros et al. \cite{BallesterosII} for different
values of $L$ and for values of $p<0.8.$ Note that the tendency of the data
for $L\longrightarrow \infty $, seems to be towards $R_{\chi }(p=0.8)$. It
means that the effect of the finite size is to introduce a apparent increase
in the value of the normalized square width which should not exist for the
sample with $L=\infty $. If we consider the data in the $p$-dependent zone,
that is $1>p>0.8$, where there is small $L$ dependence, we can make and
extrapolation to $p\longrightarrow p_{p}$ (being $p_{p}$ the probability for
which the system suffers percolation: $p_{p}\approx 0.31$\cite{Stauffer})
that is not going to be affected by finite size effects. In our case we have
used a hyperbolic tangent to fit our data, $R_{\chi }(L,1-p)=R_{\chi
}(\infty )tanh[const(1-p)]$, leaving free the universal value of the
normalized square width $R_{\chi }(\infty )$ and the slope constant ($const)$%
. Results are shown in Fig5. For $L=40,60$ we find a p-independent universal
value $R_{\chi }(\infty )\approx 0.155$, very close to previously reported
values : $0.150(7)$ \cite{BallesterosII}. For $L=20,$ on the other hand,
finite size effects are important even in the region $(1>p>0.8)$, and the
extrapolation gives a somewhat higher value $R_{\chi }(\infty )\approx 0.19$.

In conclusion, we have presented Monte Carlo data for diluted Ising systems
in the region where crossover to the diluted random universality class takes
place $(1>p>0.8).$ The evolution of the normalized square width for the
magnetization $R_{M}$ and the susceptibility $R_{\chi }$ indicates a smooth
transition with no critical probability $p_{c}$ (corresponding to a well
defined boundary between the pure and the randomly diluted universality
classes). The transition zone studied is $p$-dependent but $L$-independent.
This result is very convenient for extrapolation to the universal value $%
R_{\chi }(\infty )$ which is independent of $L.$

We acknowledge financial support from CGCyT through grant PB96-0037.

\newpage

{\bf Figure Captions}

Fig1: Critical temperature vs. spin per site probability in the slightly
diluted zone. Our data is compared with the extrapolation performed for data
by Heuer et al. \cite{Heuer} and Ballesteros et al. \cite{BallesterosII}.

Fig2: Normalized histograms for the susceptibility values obtained at
criticality. The values of the spin per site probability considered are (a)
0.8,0.825,0.85, 0.875 and (b) 0.9,0.925,0.95,0.975.

Fig3: Normalized square width for the magnetization vs. $1-p$, for values of 
$L=20,40,60$. Dotted line is just a guide for the eye.

Fig4: Normalized square width for the susceptibility vs. $1-p$, for values
of $L=20,40,60$. Dotted line is just a guide for the eye.

Fig5: Normalized square width for the susceptibility vs. $1-p$ for values of 
$L=20,40,60$ (circles), together with the data reported in \cite
{BallesterosII}. The two continuous lines indicate the universal value of $%
R_{\chi }$ for the randomly diluted Ising system reported in \cite
{BallesterosII}. The hyperbolic extrapolation of the data is indicated by a
segmented line for $L=40,60$ and by a doted line for $L=20$. \newpage 
\begin{table}[tbp]
\caption{Effective critical exponents and normalized square widths for the
pure and randomly diluted Ising system}
\label{puretable}
\begin{tabular}{cccccc}
x & $\alpha $ & $\beta $ & $\gamma $ & $R_{\chi} $ &  \\ \hline
pure (undiluted) & 0.11 & 0.3269 & 1.237 & 0.000 &  \\ 
diluted & -0.051 & 0.3546 & 1.342 & 0.155 &  \\ 
$\delta x/x$ & -1.4636 & 0.0847 & 0.0848 & $\infty$ & 
\end{tabular}
\end{table}

\newpage

\begin{figure}
\protect\epsfxsize=16cm\protect\epsfysize=20cm\protect\epsfbox{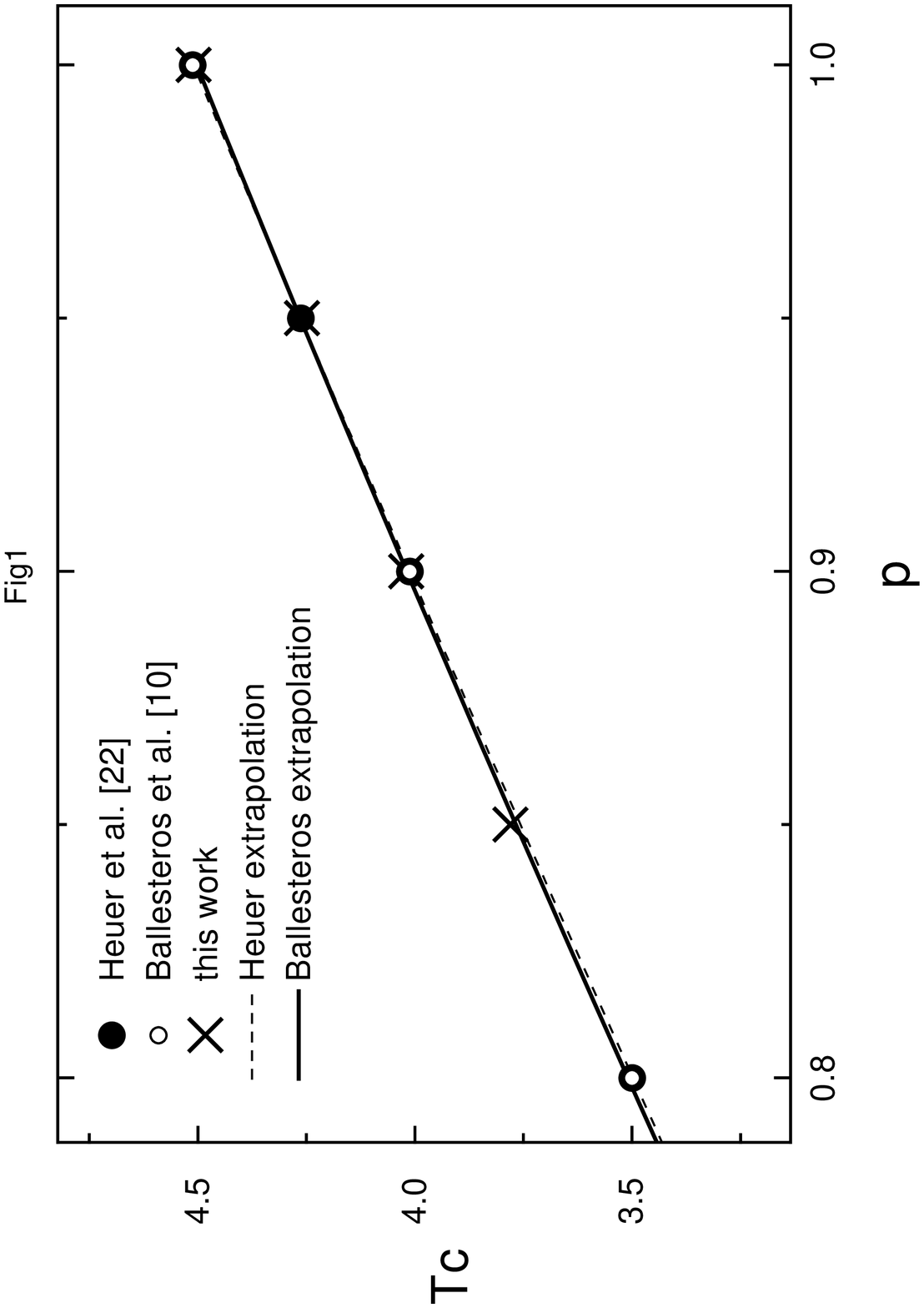}
\label{fig1}
\end{figure}
\newpage

\begin{figure}
\protect\epsfxsize=16cm\protect\epsfysize=20cm\protect\epsfbox{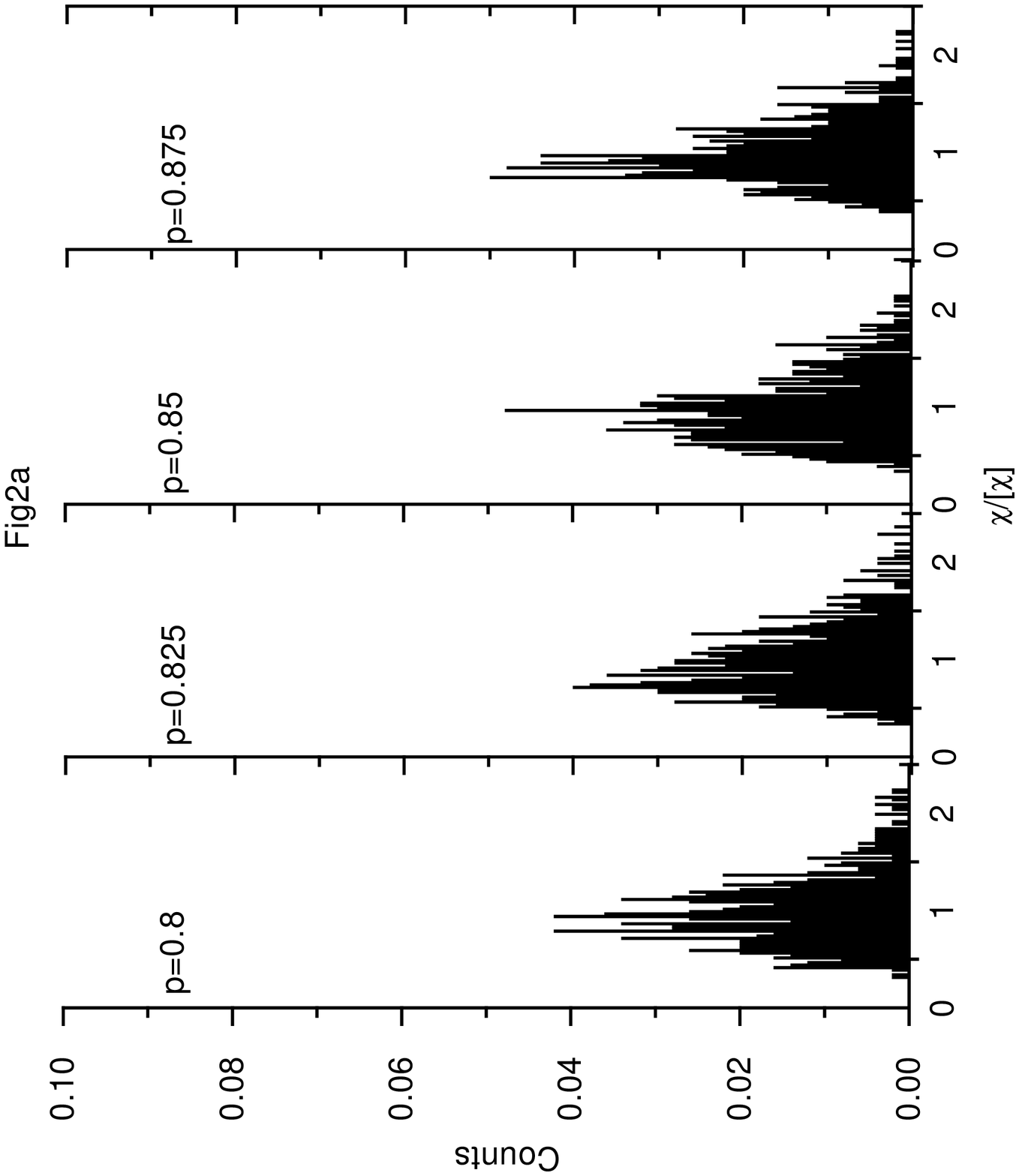}
\label{fig2a}
\end{figure}

\newpage

\begin{figure}
\protect\epsfxsize=16cm\protect\epsfysize=20cm\protect\epsfbox{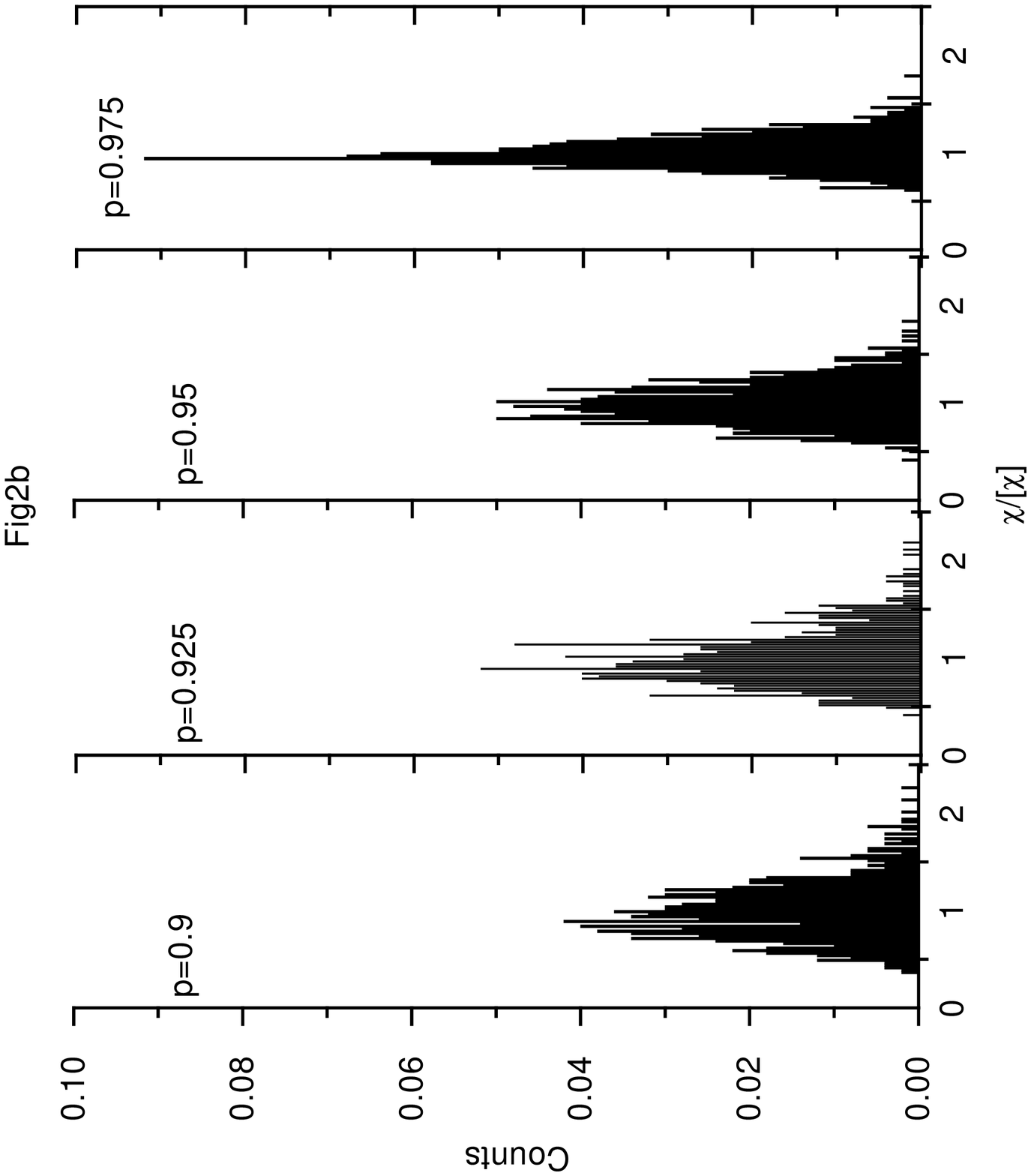}
\label{fig2b}
\end{figure}

\newpage

\begin{figure}
\protect\epsfxsize=16cm\protect\epsfysize=20cm\protect\epsfbox{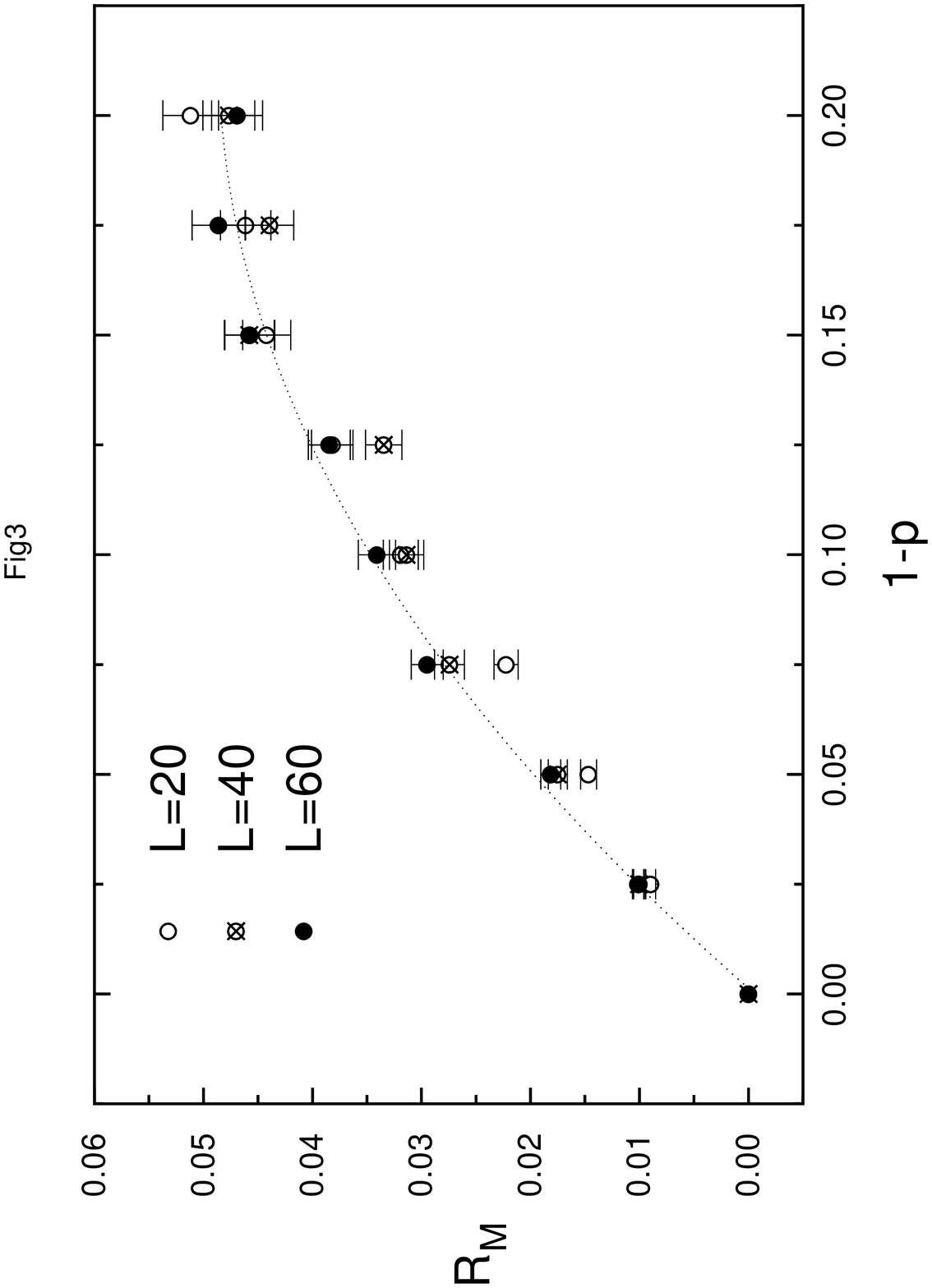}
\label{fig3}
\end{figure}

\newpage

\begin{figure}
\protect\epsfxsize=16cm\protect\epsfysize=20cm\protect\epsfbox{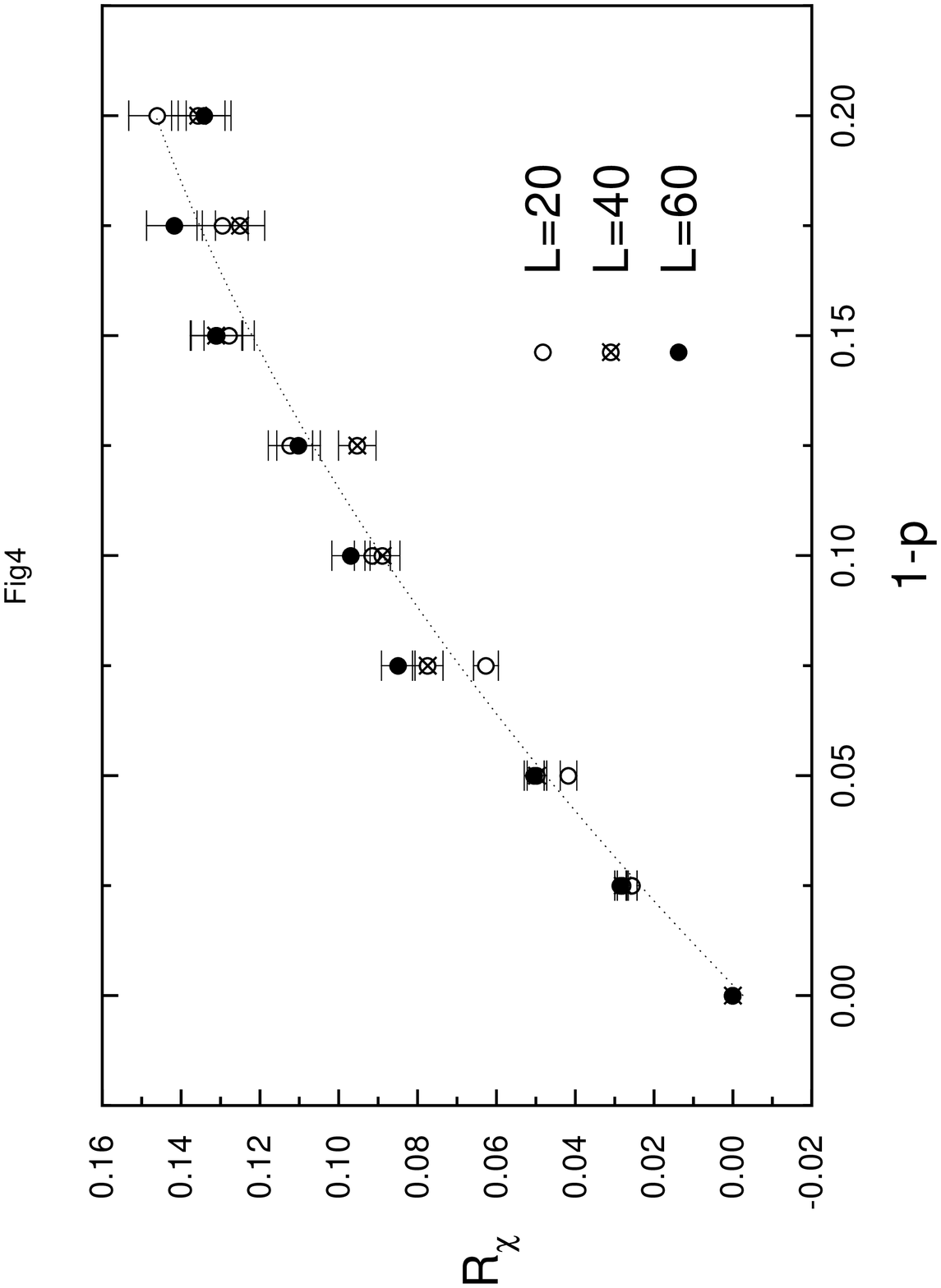}
\label{fig4}
\end{figure}

\newpage

\begin{figure}
\protect\epsfxsize=16cm\protect\epsfysize=20cm\protect\epsfbox{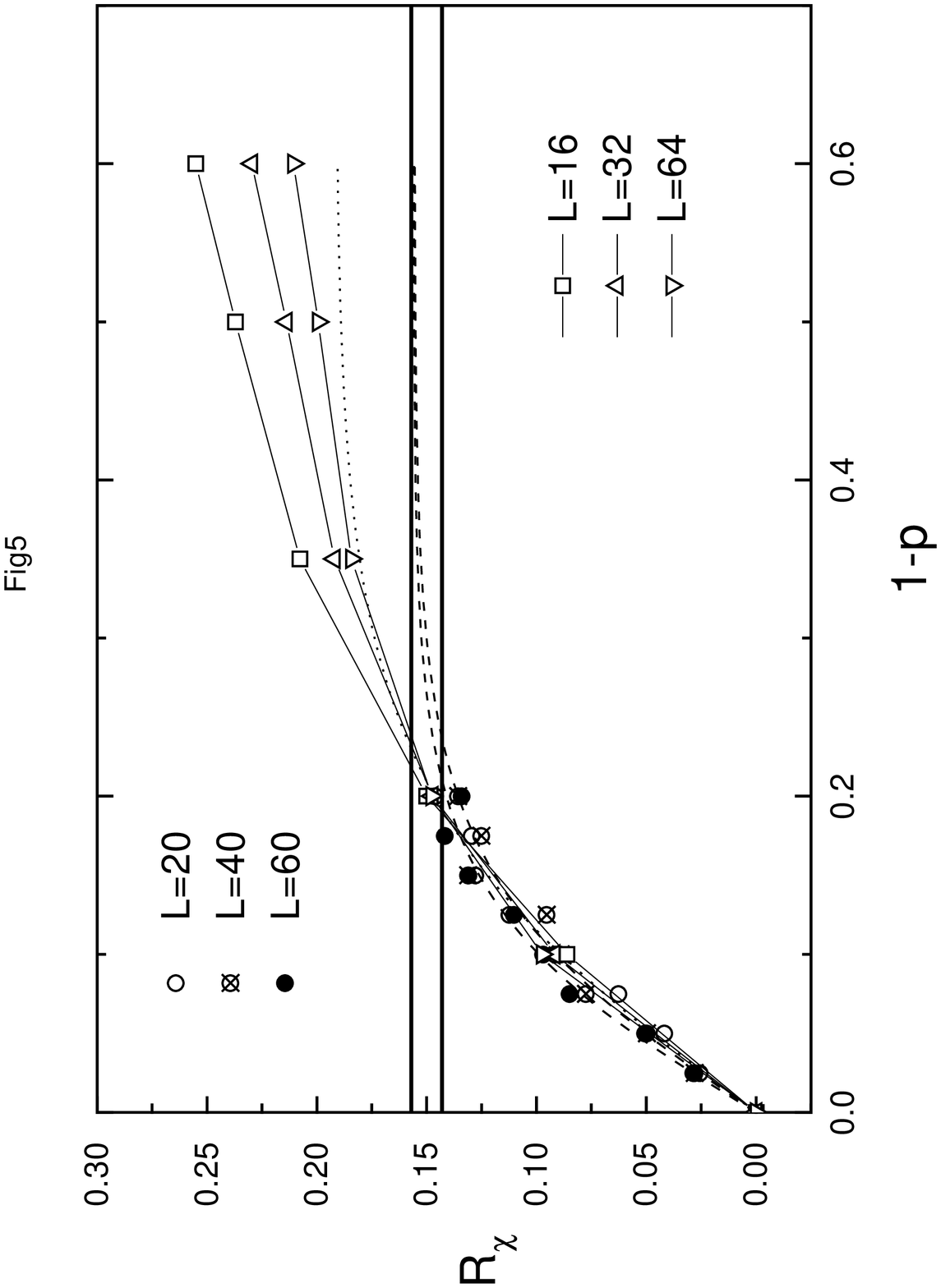}
\label{fig5}
\end{figure}

\end{document}